\documentclass[useAMS,usenatbib]{mn2e}

\usepackage{url,times,stfloats,graphicx,amsmath,amsfonts,amssymb,aas_macros,color,epsfig,epstopdf}

\newcommand{\mhalo}{M$_{\rm halo}$}
\newcommand{\mb}{M$_{\rm b}$}
\newcommand{\mstar}{M$_{\rm *}$}
\newcommand{\lcdm}{$\Lambda$CDM}
\newcommand{\msun}{M$_\odot$}

\newcommand{\vmax}{V$_{\rm max}$}
\newcommand{\vflat}{V$_{\rm flat}$}

\newcommand{\kms}{km$\,$s$^{-1}$}

\newcommand{\ahf}{\texttt{AHF}}

\newcommand{\hMpc}{{\ifmmode{h^{-1}{\rm Mpc}}\else{$h^{-1}$Mpc}\fi}}
\newcommand{\hkpc}{{\ifmmode{h^{-1}{\rm kpc}}\else{$h^{-1}$kpc}\fi}}
\newcommand{\hMsun}{{\ifmmode{h^{-1}{\rm {M_{\odot}}}}\else{$h^{-1}{\rm{M_{\odot}}}$}\fi}}
\newcommand{\ltsima}{$\; \buildrel < \over \sim \;$}
\newcommand{\gtsima}{$\; \buildrel > \over \sim \;$}
\newcommand{\lsim}{\lower.5ex\hbox{\ltsima}}
\newcommand{\gsim}{\lower.5ex\hbox{\gtsima}}

\def\lesssim{\mathrel{\hbox{\rlap{\hbox{\lower4pt\hbox{$\sim$}}}\hbox{$<$}}}}
\def\gtrsim{\mathrel{\hbox{\rlap{\hbox{\lower4pt\hbox{$\sim$}}}\hbox{$>$}}}}

\newcommand{\beq}{\begin{equation}}
\newcommand{\eeq}{\end{equation}}
\def\beqa{\begin{eqnarray}}
\def\eeqa{\end{eqnarray}}
\def\hMpc{$h^{-1}\,{\rm Mpc}$}
\def\hkpc{$h^{-1}\,{\rm kpc}$}

\def\head{
 \vbox to 0pt{\vss
                   \hbox to 0pt{\hskip 440pt\rm LA-UR-10-07069\hss}
                  \vskip 25pt}}

\title[The Mass Discrepancy--Acceleration relation]
{The distribution of mass components in simulated disc galaxies }
\author[I. Santos-Santos et al.]
       {Isabel M. Santos-Santos$^{1,2}$\thanks{E-mail: isabelm.santos@uam.es}, Chris B. Brook$^{1,2}$, 
       Greg Stinson$^{3}$, Arianna Di Cintio$^{4}$,
\newauthor    James Wadsley$^{5}$, Rosa Dom\'{\i}nguez-Tenreiro$^{1,2}$,      Stefan Gottl\"{o}ber$^{6}$,        
          Gustavo Yepes$^{1,2}$            \\
$^{1}$Departamento de F\'isica Te\'orica, Universidad Aut\'onoma de Madrid, 28049 Cantoblanco, Madrid, Spain\\
$^{2}$Astro-UAM, UAM, Unidad Asociada CSIC\\
$^{3}$Max-Planck-Institut f\"ur Astronomie, K\"onigstuhl 17,
            Heidelberg, 69117, Germany\\
$^4$Dark Fellow, Dark Cosmology Centre, NBI, University of Copenhagen, Juliane Maries Vej 30, DK-2100 Copenhagen, Denmark\\
$^5$Department of Physics \& Astronomy, McMaster University, Hamilton, 
Ontario, L8S~4M1, Canada\\
$^6$Leibniz Institute for Astrophysics Potsdam, An der Sternwarte 16, 14482 Potsdam, Germany\\
}

\begin{document}

\date{Accepted XXXX . Received XXXX; in original form XXXX}

\pagerange{\pageref{firstpage}--\pageref{lastpage}} \pubyear{2010}

\maketitle

\label{firstpage}


\begin{abstract}

Using 22 hydrodynamical  simulated galaxies in a \lcdm\ cosmological context we recover not only the observed baryonic Tully-Fisher relation,  but also the observed ``mass discrepancy--acceleration" relation, 
 which reflects the distribution of the main components of the galaxies throughout their disks.
This implies that the simulations, which span the range 52$<$\vflat$<$222\,\kms\ where \vflat\ is the  circular velocity at the flat part of the rotation curve, and match galaxy scaling relations, are able to recover the observed relations between the distributions of stars, gas and dark matter over the radial range for which we have observational rotation curve data. Furthermore, we explicitly match the observed baryonic to halo mass relation for the first time with simulated galaxies.  We discuss our results in the context of the baryon cycle that is inherent in these simulations, and with regards to the effect of baryonic processes on the distribution of dark matter. 

\end{abstract}

\noindent
\begin{keywords}
 galaxies: evolution - formation - haloes cosmology: theory - dark matter
 \end{keywords}

\section{Introduction} \label{sec:introduction}
Within a \lcdm\ context, the angular momentum of disc galaxies originates from
tidal torques imparted by surrounding structures in the expanding
Universe, prior to proto-galactic
collapse \citep{peebles69,barnes87}.  Assuming that  gas gains a similar amount of
angular momentum as the dark matter, and that
angular momentum is substantially retained as  the gas cools to the
centres of dark matter halos \citep{fall80}, then the gas will settle into a disc,
  fragment and form stars. 

Simulations  have shown that angular momentum acquisition is more complicated than this picture,  involving  a complex web structure \citep[e.g.][]{pichon11,dominguez15}. Indeed,
there has been significant progress over the past years in our ability to simulate the processes of disc formation within a cosmological context. Without an efficient feedback scheme, angular momentum is lost to dynamical friction  during the mergers of overly dense sub-structures \citep[e.g.][]{navarrosteinmetz00,maller02,piontek11}.

Progress was made by implementing increasingly effective recipes for feedback from supernovae \citep{thacker01,stinson06} and the inclusion of other forms of feedback from massive stars \citep{stinson13,hopkins14}. The benchmark for assessing this  progress has primarily been the ability to match the Tully-Fisher relation \cite[e.g.][]{governato04,domenech12}, with recent simulations succeeding at this, and in particular matching the Baryonic Tully Fisher relation (BTFR), for galaxies over a range of masses \citep{brook12,aumer13}.

\begin{table*}
\begin{center}
\caption{Properties of the simulated galaxies ordered by halo mass. MaGICC galaxies have a "g" as prefix, while CLUES galaxies have a "C".
Disk scale lengths $h$ and central surface brightnesses $\mu_0$ are derived from exponential fits to the surface brightness profile in the I band. }
\begin{tabular}{l l l l l l l l}
\hline
Name &  \mhalo\ (\msun) & \mstar\ (\msun) & M$_{\rm HI}$ (\msun) & $h$ (kpc) & $\mu_0$ (mag as$^{-1}$) & \vmax\ (\kms)& V$_{\rm flat}$ (\kms)  \\
\hline
\hline
g15784\_MW &1.49$\times$10$^{12}$& 5.67$\times$10$^{10}$& 1.96$\times$10$^{10}$    &3.23 & 19.09 &  222.10  & 222.10 \\
g21647\_MW  &8.24$\times$10$^{11}$& 2.51$\times$10$^{10}$& 5.62$\times$10$^9$    & 1.30 &  17.39 &   189.79 &  163.84 \\
g1536\_MW  & 7.06$\times$10$^{11}$&2.36$\times$10$^{10}$& 6.78$\times$10$^9$   &3.46 &   20.54 & 175.95 & 175.95 \\
g5664\_MW  & 5.39$\times$10$^{11}$&2.74$\times$10$^{10}$& 4.19$\times$10$^9$   &2.34 &   19.51 &  196.66& 151.40 \\
g7124\_MW  & 4.47$\times$10$^{11}$&6.30$\times$10$^9$& 3.49$\times$10$^9$   &2.79 &  20.60 &  120.14  &  120.14\\
g15807\_Irr &2.82$\times$10$^{11}$& 1.46$\times$10$^{10}$& 4.68$\times$10$^9$  &1.94 &  18.68 & 141.21 & 141.21 \\
g15784\_Irr &1.70$\times$10$^{11}$& 4.26$\times$10$^9$& 2.70$\times$10$^9$  &2.27 & 20.30 &  106.90 &  106.90\\
g22437\_Irr  &  1.10$\times$10$^{11}$&7.44$\times$10$^8$& 1.08$\times$10$^9$  &1.88 &  21.33 &  75.40 &  75.40 \\
g21647\_Irr  &  9.65$\times$10$^{10}$&1.98$\times$10$^8$&  3.68$\times$10$^8$  &1.75 & 22.77 &  60.85&  60.85 \\
g1536\_Irr & 8.04$\times$10$^{10}$&4.46$\times$10$^8$& 4.39$\times$10$^8$   &1.70 &  21.75 &  67.16 &  67.16\\
g5664\_Irr  &  5.87$\times$10$^{10}$& 2.36$\times$10$^8$&  2.56$\times$10$^8$   &1.66 &   22.28 &  59.50 &  59.50 \\
g7124\_Irr  & 5.23$\times$10$^{10}$&1.32$\times$10$^8$& 2.30$\times$10$^8$   &1.16 &   21.61 & 52.77& 52.77 \\
\hline
C1 & 7.23$\times$10$^{11}$ & 1.45$\times$10$^{10}$ &  3.86$\times$10$^9$ &  1.56 & 19.56 &168.83 & 127.10 \\
C2 & 5.31$\times$10$^{11}$ & 1.11$\times$10$^{10}$ &  6.32$\times$10$^8$ &  1.83 & 20.40 & 123.59 & 123.59\\
C3 & 2.67$\times$10$^{11}$ & 5.08$\times$10$^9$ &  2.79$\times$10$^9$ &  2.25 & 21.22 & 119.75 &119.75\\
C4 & 1.87$\times$10$^{11}$ & 4.18$\times$10$^9$ &  9.77$\times$10$^7$ &  1.45 & 20.07 & 101.21 & 101.21\\
C5 & 1.51$\times$10$^{11}$ & 4.54$\times$10$^9$ &  2.42$\times$10$^9$ &  1.35 & 20.30 & 116.63 & 116.63 \\
C6 & 1.29$\times$10$^{11}$ & 2.08$\times$10$^9$ &  2.74$\times$10$^9$ &  1.53 & 21.29 & 101.66 & 101.66\\
C7 & 1.18$\times$10$^{11}$ & 1.57$\times$10$^9$ &  9.10$\times$10$^8$ &  1.48 & 20.44 & 88.89 & 72.92 \\
C8 & 1.21$\times$10$^{11}$ & 1.57$\times$10$^9$ &  6.34$\times$10$^8$ &  1.03 & 20.19 & 85.22 & 85.22 \\
C9 & 8.04$\times$10$^{10}$ & 1.10$\times$10$^9$ &  1.05$\times$10$^8$ &   1.55 & 22.64 & 70.85 & 70.85\\
C10 & 6.44$\times$10$^{10}$ &3.78$\times$10$^8$ &  6.33$\times$10$^7$ &  0.86 & 21.55 &  53.35 & 53.35 \\
\hline
\end{tabular}
\label{tab:sims}
\end{center}
\end{table*}

Yet rotation curves of observed galaxies provide significantly more information regarding the angular momentum of galaxies than is contained within the BTFR, allowing  more stringent  constraints on galaxy formation models which have not previously been applied to simulated galaxies.
 High resolution observations  of HI velocities,  combined with  studies of the gas and stellar mass distributions, provide detailed information on how the different  mass components are radially  distributed in galaxies with a wide range of rotational velocities V$_r$ \cite[e.g.][]{begeman91,sanders98,deblok01,gentile04,kuziodenaray06,oh15}.  
 
 Differences between the mass implied by measured rotational velocities, and the baryonic  mass observed in the form of gas and stars, is usually attributed to dark matter \citep{rubin70}, an assumption which our simulations embrace. In this study we aim to determine whether  galaxies simulated in a \lcdm\ universe  can reproduce the detailed radial mass distribution of observed galaxies.  
 
 An instructive way to display the radial mass distribution of a population of disc galaxies is to plot the mass discrepancy-acceleration relation \citep{sanders02,mcgaugh04}. Mass discrepancy, D, is defined as the  ratio of the square of the measured rotation velocity, and the square of the rotation velocity that can be attributed to baryons, D$\equiv$(V$_r$/V$_b$)$^2$. The acceleration is  defined at each radius, r, by the baryonic contribution to the  centripetal acceleration, g$_b$$\equiv$V$_b$$^2$/r. 

Despite a large variety in rotation curve shapes \citep[e.g.][]{zwaan95,tully97,swaters09}, disc galaxies with a wide range of V$_r$ show a remarkably tight D-g$_b$ relation \citep{mcgaugh04,mcgaugh14}.  Galaxies that present the same mass discrepancy at various  radii  all experience, at those radii,  the same gravitational radial force as contributed by their baryons; it is as though the rotation velocity attributed to dark matter depends only on the distribution of baryonic mass. Indeed, this tight relation has been interpreted as being causal  \citep{sanders02,mcgaugh14}, and therefore  evidence for modified Newtonian dynamics, MOND \citep{milgrom83}.   This empirical result  has no {\it a priori} explanation in a $\Lambda$CDM cosmology, but a study of semi-analytic  models \citep{vdb00} found that galaxies tuned to match the Tully-Fisher relation reveal a characteristic acceleration.  



Here,  we explore the D-g$_b$ relation in a suite of 22  disk galaxies  simulated within a \lcdm\ context, which vary in their virial, stellar and baryonic masses (\mhalo, \mstar, \mb), star formation histories, disk scale lengths, central surface brightnesses and circular velocity curve shapes. As we show, the suite of galaxies tightly  match the empirical \mstar-\mhalo\ \citep{guo10,moster10}, \mb-\mhalo\ \citep{papastergis12} and BTFRs \citep{mcgaugh05}. 

It is important to emphasise that these simulated galaxies were not tuned  to reproduce the BTFR, with free parameters tuned to match the stellar to halo mass relation at one halo mass \citep{brook12,stinson13}, and  then fixed in simulations of different mass halos. The simulations  have  previously been shown to match a wide range of scaling relations including the Tully-Fisher, luminosity-size, mass-metallicity relations, and  HI mass to r band luminosity ratio as a function of R band magnitude,  at z=0 \citep{brook12}. Further, the simulations  match the 
evolution of the stellar mass-halo mass relation 
\citep{stinson13,kannan13}, as derived by abundance matching 
\citep{moster13} and a range of relations at high redshift \citep{obreja14}. The simulations also
expel sufficient metals to match local observations 
\citep{prochaska11,tumlinson11} of OVI in the circum-galactic medium 
\citep{stinson12,brook12}.

The paper is organized as follows. Section 2 presents the simulations used, describing their initial conditions and  baryonic modelling. The circular velocity curves, BTFR, \mstar-\mhalo\, \mb-\mhalo\ relations and plots of D versus g and  radius are shown in Section 3. Residuals around the D-g relation are also shown. 
Finally, Section 4 discusses the implications of our results.


\section{The Simulations}\label{simulations}

\begin{figure*}
\hspace{0.cm}\includegraphics[width=\textwidth]{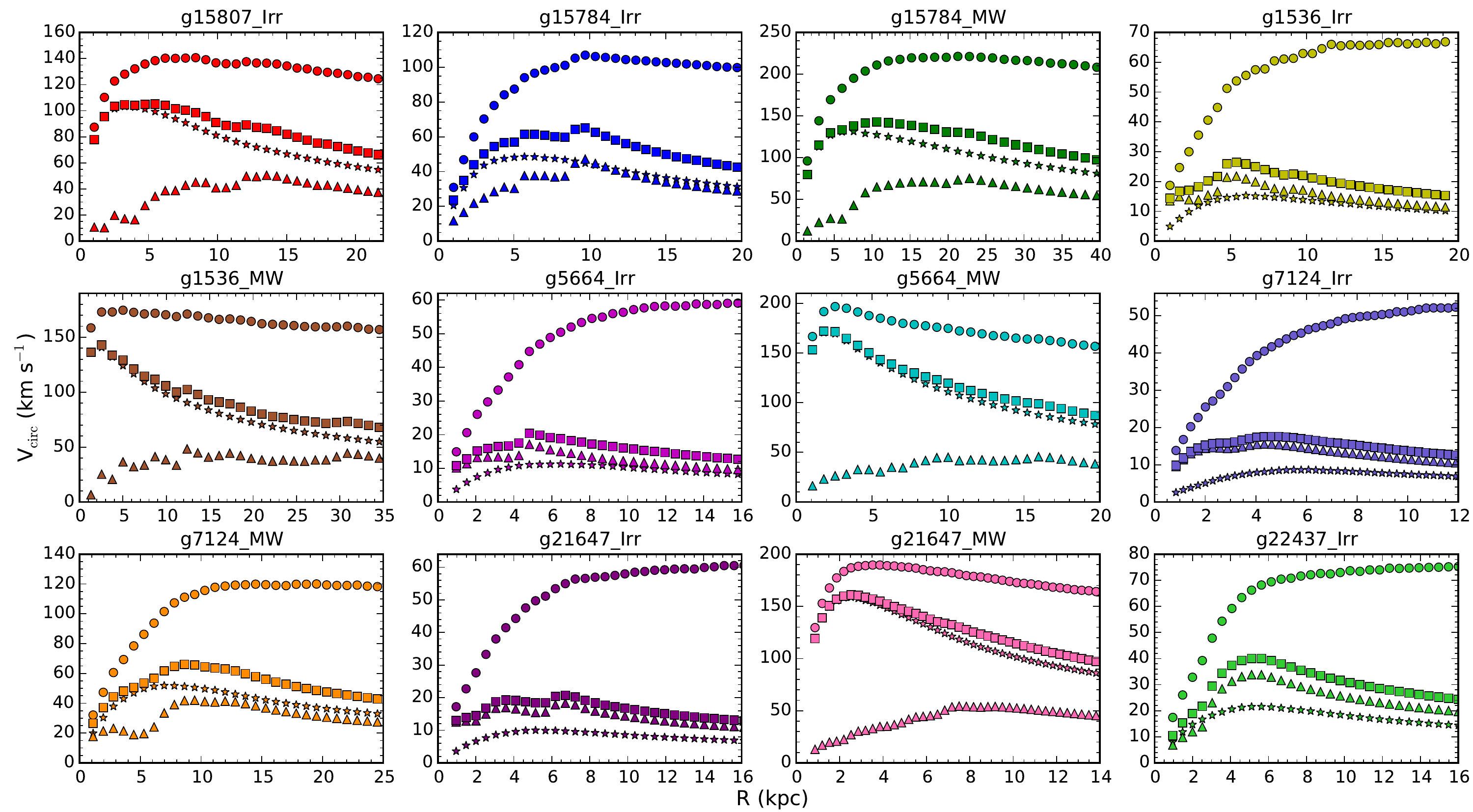}
\caption{The circular velocity curves of the 12 MaGICC disk galaxies. Different symbols represent the velocity values due to different mass components (triangles: cold gas; stars: stars; squares: all baryons;
 circles: total). Simulations reproduce the variety of observed rotation curves. Furthermore, like in observations, the features present in the baryonic curves are reflected in the total one.
}
\label{fig:magicc_rc}
\end{figure*}

Two sets of simulated galaxies have been used, with slightly different input physics, as we will explain. In first place, we use 12 galaxies from the MaGICC (Making Galaxies in a Cosmological Context) project \citep{brook12a,stinson13}. These are zoomed-in regions of a total cosmological volume of side 68 Mpc. 
 The resolution varies depending on the ``type" of simulated galaxy, labelled MilkyWay or Irregular. The former have m$_{\rm star}$=4.0$\times$10$^4$M$_{\odot}$, m$_{\rm gas}$=5.7$\times$10$^4$M$_{\odot}$, m$_{\rm dm}$=1.1$\times$10$^6$M$_{\odot}$ and a gravitational softening length of $\epsilon$=312pc (for all particle types),
 while the latter are more highly resolved with m$_{\rm star}$=4.3$\times$10$^3$M$_{\odot}$, m$_{\rm gas}$=7.1$\times$10$^3$M$_{\odot}$, m$_{\rm dm}$=1.4$\times$10$^5$M$_{\odot}$ and $\epsilon$=156pc.
The initial 
power spectrum of density fluctuations is derived from the McMaster Unbiased Galaxy Simulations (MUGS) \citep{stinson10} which use a \lcdm\ cosmology with WMAP3 parameters, i.e. $H_0=73$ \kms\ Mpc$^{-1}$, $\Omega_{\rm m}=0.24$, $\Omega_{\Lambda}=0.76$, $\Omega_{\rm baryon}=0.04$ and $\sigma_8=0.76$.

The second set of galaxies is from a single simulation with initial conditions from the CLUES project (Constrained Local UniversE Simulations, \citealt{gottloeber10,yepes14}).
 Again the zoom-in technique is used, this time together with observational data (masses of nearby X-ray clusters and peculiar velocities obtained from catalogs) imposed as constraints on the initial conditions, in order to simulate a cosmological volume that is representative of our local universe.
The Hoffman-Ribak algorithm, using the observational data mentioned above, is used to constrain scales down to $\approx 5h^{-1}$Mpc. This way structures like the Virgo cluster, Coma cluster and Great attractor, are always reproduced by the simulations.
 
 On  smaller scales,  the distribution of structure is essentially random, and
  several dark matter-only realizations are run until a  Local Group analogue (a Milky Way-M31 like binary group) is found. Then this Local Group region is re-simulated with baryons and at a higher resolution. In this case, the re-simulation includes 4096$^3$ effective particles in a spherical volume of 2$h^{-1}$Mpc around the Local Group. The mass resolution of particles is m$_{\rm star}$=1.3$\times$10$^4$M$_{\odot}$, m$_{\rm gas}$=1.8$\times$10$^4$M$_{\odot}$ and m$_{\rm dm}$=2.9$\times$10$^5$M$_{\odot}$, and the gravitational softening lengths are $\epsilon_{\rm bar}$=223pc between baryons and $\epsilon_{\rm dm}$=486pc between dark matter particles.
This CLUES simulation also follows a WMAP3 cosmological model.

The CLUES simulation used  is not one of the previously published CLUES simulations, which were evolved using the PMTree-SPH MPI code GADGET2, but rather part of a new set with the same initial conditions but different physics prescriptions for star formation and feedback, as explained below.

All simulations in this study are evolved using the parallel N-body+SPH tree-code \verb,GASOLINE, \citep{wadsley04}, which includes gas hydrodynamics and cooling, star formation, energy feedback and metal enrichment to model structure formation. We describe here the most important implementations (for details see \citealt{governato10} and \citealt{stinson13}).

When gas gets cold and dense, stars are formed according to a Schmidt law with star formation rate $\propto\rho^{1.5}$. Stars feed energy and metals to the surrounding interstellar medium. Energy feedback by supernovae is implemented by means of the blastwave formalism \citep{stinson06} where $\epsilon_{\rm SN}\times 10^{51}$ erg of thermal energy is released.
The amount of metals deposited from SNe explosions is computed from a Chabrier IMF, and they diffuse between gas particles as described in \citet{shen10}. \verb,GASOLINE, also accounts for the effect of a uniform background radiation field on the ionization and excitation state of the gas.
In the case of the MaGICC simulations, metal-line cooling \citep{shen10} and early stellar feedback from massive stars \citep{stinson13} prior to their explosion as SNe are also included.

\begin{figure*}
\includegraphics[width=1.\textwidth]{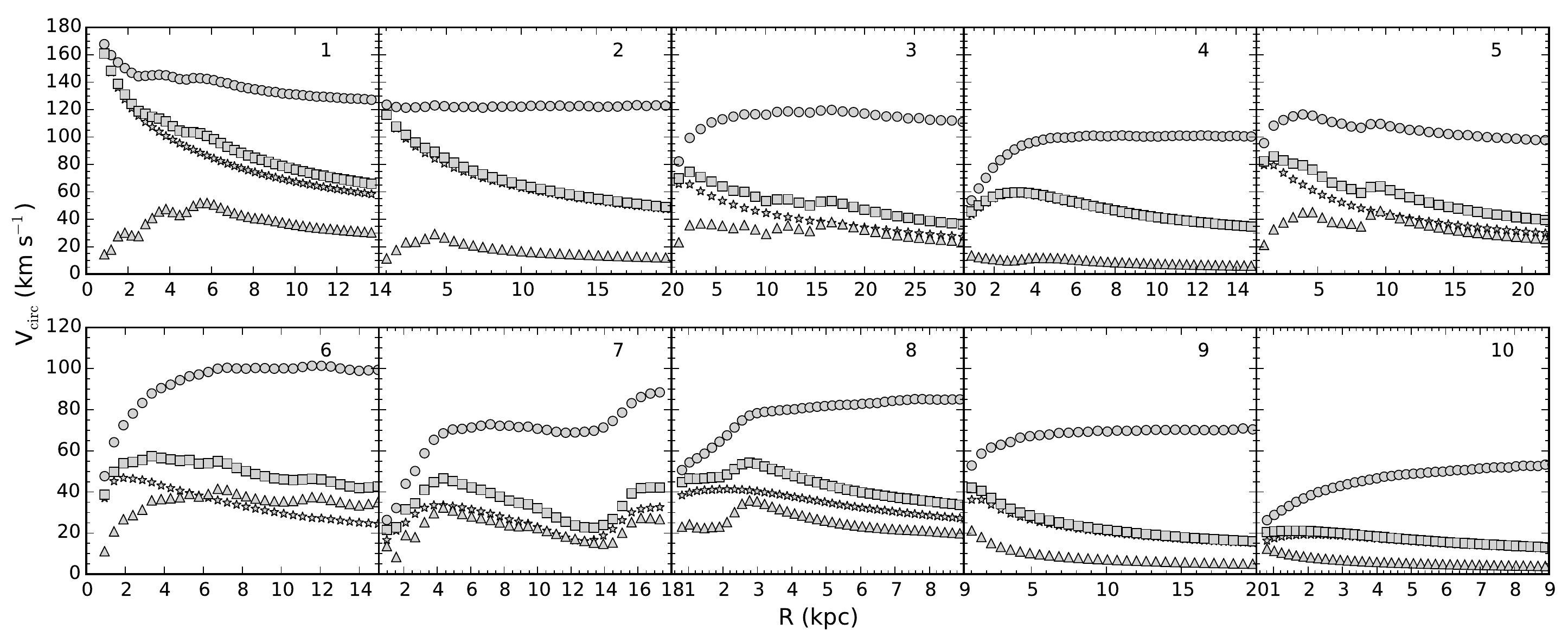}
\caption{The circular velocity curves of the 10 galaxies selected from the CLUES simulation. 
Triangles: cold gas; stars: stars; squares: all baryons; 
circles: total.} 
\label{fig:clues_rc}
\end{figure*}

The CLUES simulations 
follow  the physics used in \citet{governato10} and \citet{guedes11}, which formed   realistic dwarf and Milky Way galaxies respectively. These runs do not include metal line cooling, nor do they include early stellar feedback. As argued in \citet{feldman15}, gas cooling is also affected by UV and soft X-ray emission from nearby massive stars (e.g. \citealt{cantalupo10,Kannan14b}); it is not clear whether adding metal line cooling, without their potential counterparts, such as radiative ionization by local sources, results in a better model. 

Further, in these types of simulations, feedback recipes are not well constrained, but are basically tuned to balance whatever cooling rate is included, in order to match the constraints imposed; in these simulations,   constraints come from the stellar  to halo mass relation.    
The lower cooling rates of the CLUES set of simulations is compensated by the lack of early stellar feedback, resulting in our two sets of simulations following very similar trends in their structural properties, as we will see below. One could argue that sub grid local feedback processes are included in the CLUES simulation by adjusting the cooling function. 

 We emphasise that the two sets of simulations share the same  implementations of SNe feedback and star formation. Yet there are differences in cooling and feedback, which will result in differences in the amount of gas cycling through central regions of the galaxies. We will comment on some systematic differences between the two sets of simulations, in terms of the relations explored in this paper.

Halos in both simulations have been identified using Amiga's Halo Finder (\ahf; \citealt{knollmann09}), where their masses are defined as the mass inside a sphere containing $\Delta_{\rm vir}\simeq$350 times the cosmic background matter density at redshift z=0. 

The analysis of the simulation data was largely performed using the open source PYNBODY package \citep{pynbody}.


The  properties of the simulated galaxies are presented in Table \ref{tab:sims}. The MaGICC simulations are 12 disk galaxies, separated into two sub-sets labelled as Milky-Way (MW) and irregular (Irr) type galaxies, although they all are disc galaxies with stellar masses ranging from 1$\times$10$^{8}$-5$\times$10$^{10}$\msun. 
From the CLUES simulation we have selected the halos that satisfy the following conditions: (i) Not a sub-halo, (ii) \mhalo$>$4$\times$10$^{10}$\msun. These integrate a sample of 10 well resolved isolated galaxies. Since this is a Local Group simulation, the three most massive galaxies are loose analogues of the Milky Way, M31 and M33, and the rest are isolated dwarf galaxies.

\section{Results}  
We emphasise the MaGICC set of simulations by showing each individual galaxy in colour in all plots. This is because these galaxies have been thoroughly explored in the literature, as noted in the introduction. The CLUES simulations have  not been as extensively analysed in other contexts, and are shown  as grey dots. Considering that our results emphasise the ability of the suite of simulations to match various relations,  and that one may expect any differences in the two sets of simulations  to increase any scatter found around the relations we explore,  we feel that it is justified to  include all galaxies in the derived results. Thus, our fits  include all simulated galaxies. Nothing in our conclusions changes if only MaGICC galaxies are included, although the number and diversity of galaxies would be less.

\subsection{Circular velocity curves}\label{rcs}

Figures \ref{fig:magicc_rc} and \ref{fig:clues_rc} show  the gaseous (triangles), stellar (stars), baryonic (squares) and total (circles) circular velocity curves of the MaGICC and CLUES simulated galaxies, respectively. These are measured at radii ranging from 0.7kpc to 10$\times h$ where $h$ is the disc scale length (see Table \ref{tab:sims}). 
These circular velocities are calculated using the gravitational potential  along the midplane of the aligned simulated disc. We checked that our  results are not significantly changed by assuming a spherical potential and  averaging the mass within spherical shells (i.e., the classical GM/r potential).

 The simulated galaxies reach a flat value of circular velocity which persists to large radii, and  lack the strong peak at small radii that not so long ago was ubiquitous in simulations due to overcooling. A couple of galaxies, g5664\_MW and C1, do have significant bulges, which is reflected in  the heightened  inner region of their circular velocity curves.

The  differences in the physical modelling and initial conditions used in both simulations are not readily visible, with galaxies of similar masses reaching similar maximum, flat velocities (see for example g7124\_MW \& C3, g15807\_Irr \& C1 or g5664\_Irr \& C10).
However, there does appear to be  a tendency for the MaGICC galaxies to have more slowly rising rotation curves than the CLUES simulations. The broad range of observed rotation curve shapes has been noted for some time  \citep[e.g.][]{zwaan95,swaters09}, with recent attempts to quantify this range \citep{oman15,brook15a} and compare with cosmological models. The relatively small number of galaxies of each different suite used in this study means it is difficult to compare with observations in  a quantitative manner.
Whether  a larger suite of cosmological simulations can match the  range of observed rotation curve shapes will be explored in a later study (Santos-Santos et al. in prep). For this study, we note that there may be slight systematic  differences between  the MaGICC and CLUES rotation curves shapes, which we will explore  in terms of the mass discrepancy relation, and will be seen to be relatively minor. 
 
 Visually, one can appreciate that overall, the simulations do produce diversity in rotation curve shapes, and that the baryon contribution increases with increasing mass. One can also see that features from the baryonic components are often reflected in the total-components curves. This is known as "Renzo's Rule" \citep{renzo04,mcgaugh14}, and has long been observed in real galaxies.  In particular, these bumps and features are  noticeable in galaxies g15807\_Irr, g15784\_Irr, g1536\_MW, g5664\_MW of Figure \ref{fig:magicc_rc}, and C1, C5, C6, C7 \& C8 of Figure \ref{fig:clues_rc}. 
These results represent  evidence that the different mass components affect each other throughout the disc region as they co-evolve within a \lcdm\ Universe.

\begin{figure}
\includegraphics[width=\linewidth]{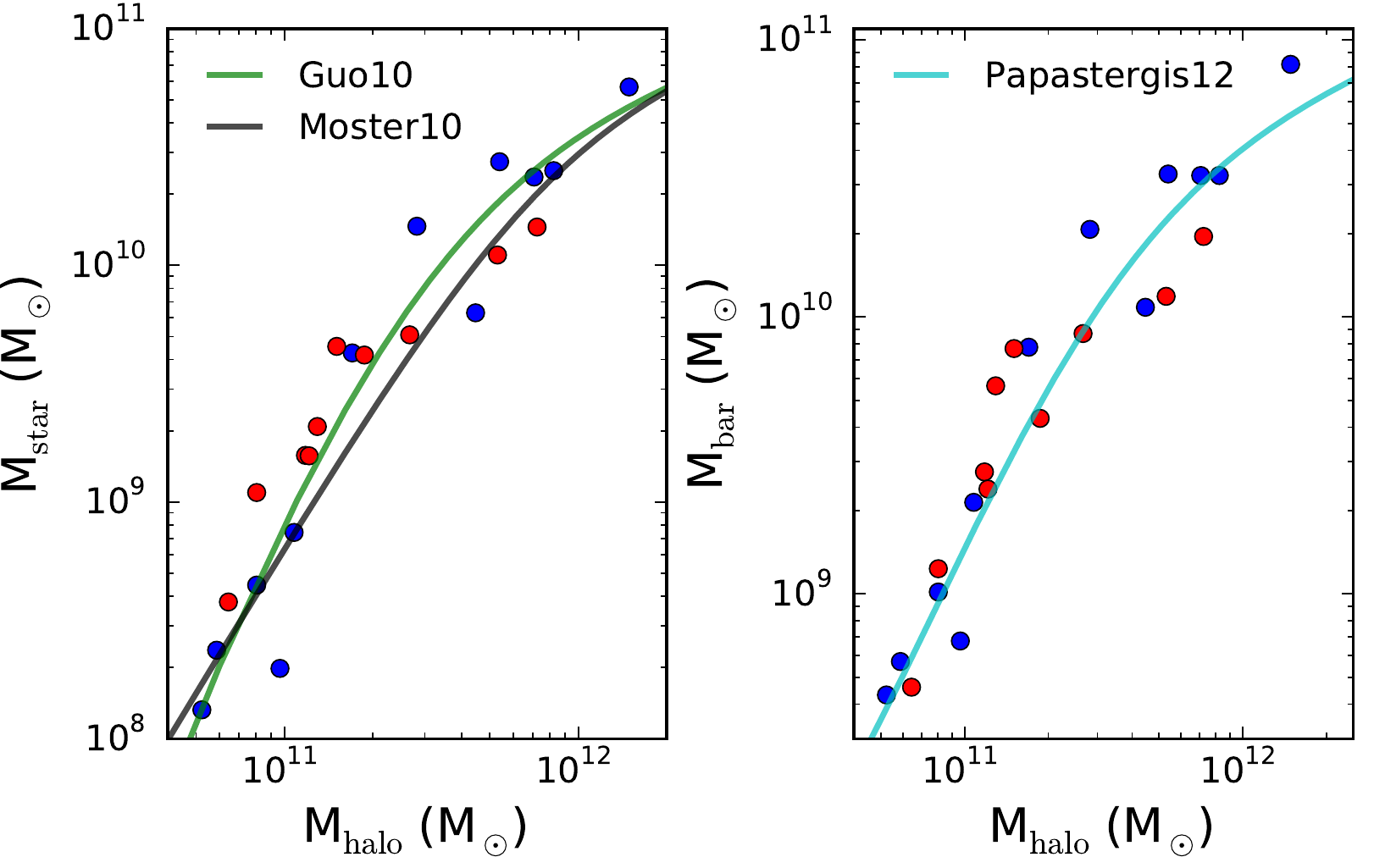}
\caption{The stellar-to-halo mass (left panel) and baryon-to-halo mass (right panel) relations, with MaGICC galaxies in blue and CLUES galaxies in red. Also shown are the empirical stellar-halo mass relations from Guo et al. 2010 (green line) and Moster et al. 2010 (black line), and the baryon-to-halo mass relation from Papastergis 2012 (cyan line).}
 \label{fig:msmh}
\end{figure}

\begin{figure}
\includegraphics[width=.45\textwidth]{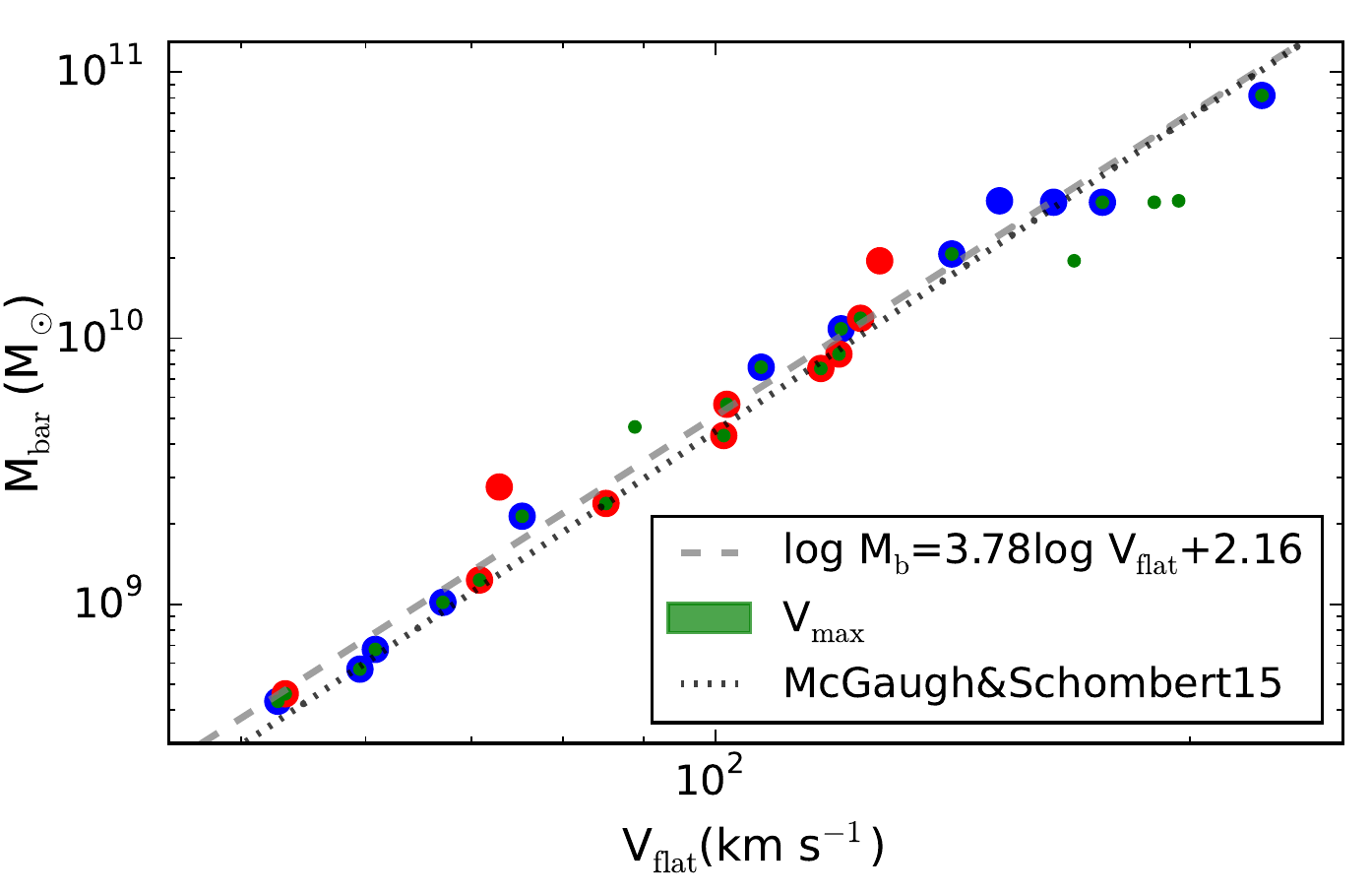}
\caption{The baryonic Tully-Fisher relation: total baryonic mass M$_{\rm bar}$ (stars + cold gas) plotted against rotation velocity \vflat. Blue points are from the MagiCC suite while the red points are the galaxies from the CLUES simulation. The dashed line shows the linear fit to the simulated data, with slope=3.78. The small green points show \vmax\ rather than \vflat, which results in a slightly flatter relation, with slope=3.49 (see text for details).
The dotted line is the observational relation using measurements in the V band given in \citealt{McGaugh15} with slope=3.92.
}
 \label{fig:btf}
\end{figure}

\subsection{Baryonic and Halo Masses}
In the left panel of  Figure \ref{fig:msmh} the stellar-to-halo mass relation of the simulations is plotted, along with the empirical relation as determined by \cite{guo10} and \cite{moster10},  whilst in the right panel of  Figure \ref{fig:msmh} the baryon-to-halo mass relation is plotted, along with the empirical relation as determined by \cite{papastergis12}. The baryonic mass is defined as the sum of the mass coming from stars and cold gas particles, where the latter is estimated as a multiple of the atomic HI gas mass M$_g=\eta$M$_{HI}$, with $\eta=4/3$ (following e.g. \citealt{mcgaugh12}).
 The Saha equation is solved to determine an ionization equilibrium and the HI mass. This remains an approximation since an accurate model of HI mass would require full radiative transfer. In particular self shielding from the UV background is not included in our model and may affect our derived HI masses, while photo-ionization of HI from the galaxy itself is also excluded. 

As stated in Section~\ref{simulations}, the MaGICC simulations were tuned to match the stellar mass-halo mass relation at one galaxy mass (in particular, to match the stellar to halo mass of galaxy g15784\_Irr), and shown to then match the relation over a range of masses (\citealt{brook12,obreja14}, see also the Nihao simulations, \citealt{wang15}, which use very similar implementation of physics, and note that other models are also able to match the relation, e.g. \citealt{munshi13,schaye15}). The CLUES simulations were also calibrated to match the relation. So, although in some sense it is not surprising that the simulations match the relation to which they were tuned, they actually match the relation over a far wider mass range than the one on which the parameter search  was performed. 

We show that the simulations also match the empirical \mb-\mhalo\ relation,  implying that they also have the same total amount of cold gas as observed galaxies at $z=0.$ As far as we know, this is the first time that simulations have been shown to match this important empirical relation, and emphasise that it was not a direct result of a parameter search. 

\subsection{The Baryonic Tully-Fisher relation}
The maximum velocity found in each simulated galaxy, \vmax, is a good approximation of the  flat velocity, \vflat, in most cases. For the cases mentioned above in which a couple of MW type galaxies have significant bulges, we show different values of \vmax\ and \vflat\ in Table~\ref{tab:sims}. 

In Figure~\ref{fig:btf} we plot the BTFR using \vflat, with the MaGICC and CLUES sets of simulations shown as blue and red dots, respectively.   In the case of C7, the galaxy is about to undergo a merger, and we use the maximum velocity from the inner 10\,kpc as \vflat\ which is the central galaxy, and use the baryonic mass from within this same radius.
 The fit to the \vflat\ BTFR  is 
 $$
\rm log \rm M_{\rm b} =3.78 \rm log \rm V_{\rm flat} + 2.16
$$
The scatter is very small, with the galaxy that is furthest from the fit being C7, the one which has a very close companion galaxy with which it is dynamically interacting. 
 
If we simply use \vmax\ in each case,  the relation is slightly flatter, and can be seen as small green dots in Figure~\ref{fig:btf}, with fit
$$
\rm log \rm M_{\rm b} =3.49 \rm log \rm V_{\rm max} + 2.67
$$
and similar scatter.

These fits are  consistent with the observational fits found in the literature \citep[see][for a summary]{mcgaugh12}, as is the trend for a flatter relation
 when using \vmax\ rather than \vflat.

\begin{figure}
\includegraphics[width=\linewidth]{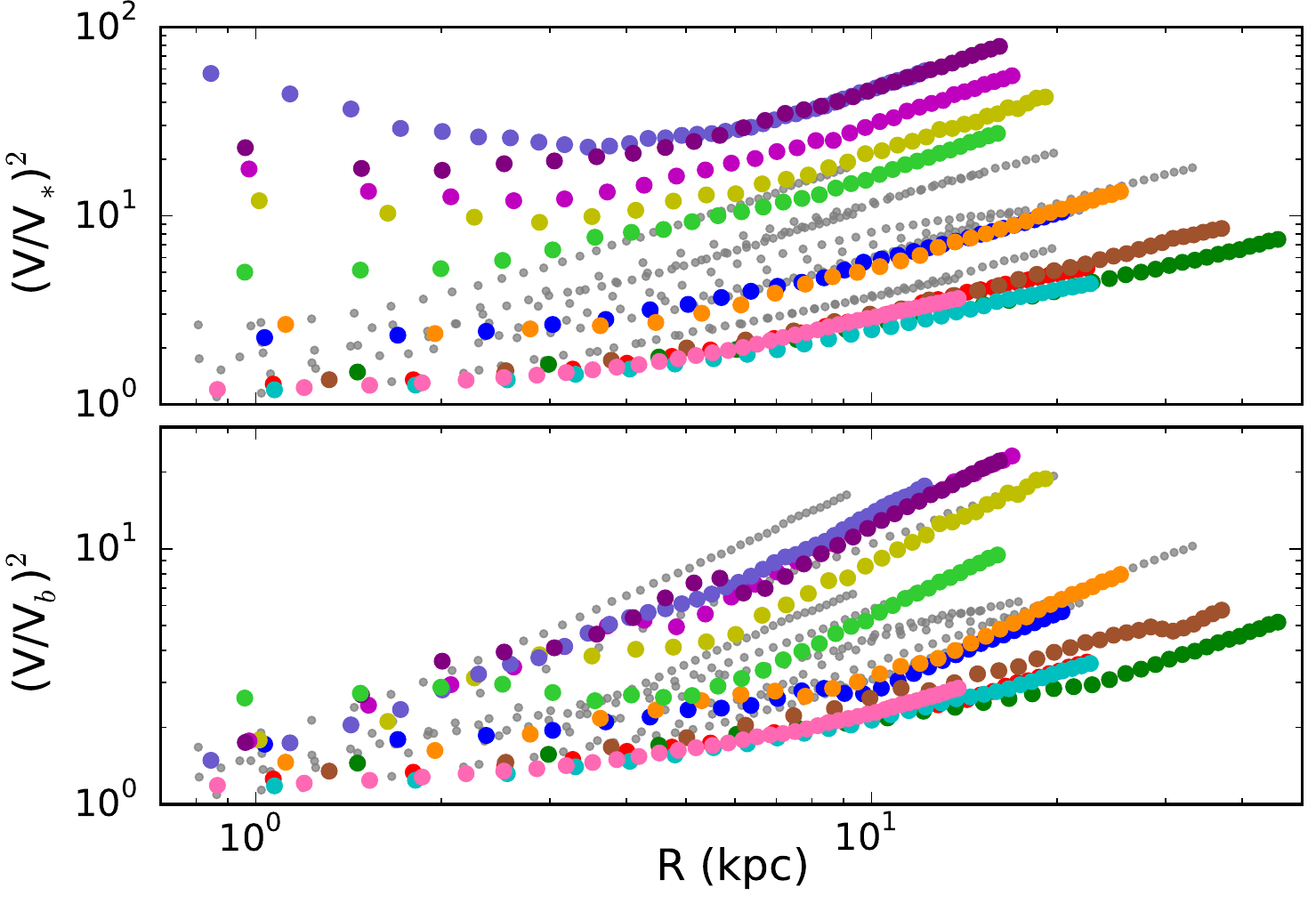}
\caption{Mass discrepancy versus radius (stars, top panel; all baryons, lower panel). Data points for each MaGICC galaxy are represented in a different color according to figure \ref{fig:magicc_rc}. Data points for all CLUES galaxies are small dots in gray color (as in figure \ref{fig:clues_rc}).
 As occurs with observed galaxies, the lower the surface brightness of the galaxy the higher the mass discrepancy encountered: mass discrepancy does not hold a correlation with radius. Smaller values of D are found when all baryons are taken into account as expected. }
\label{fig:DvsR}
\end{figure}

\begin{figure}
\includegraphics[width=\linewidth]{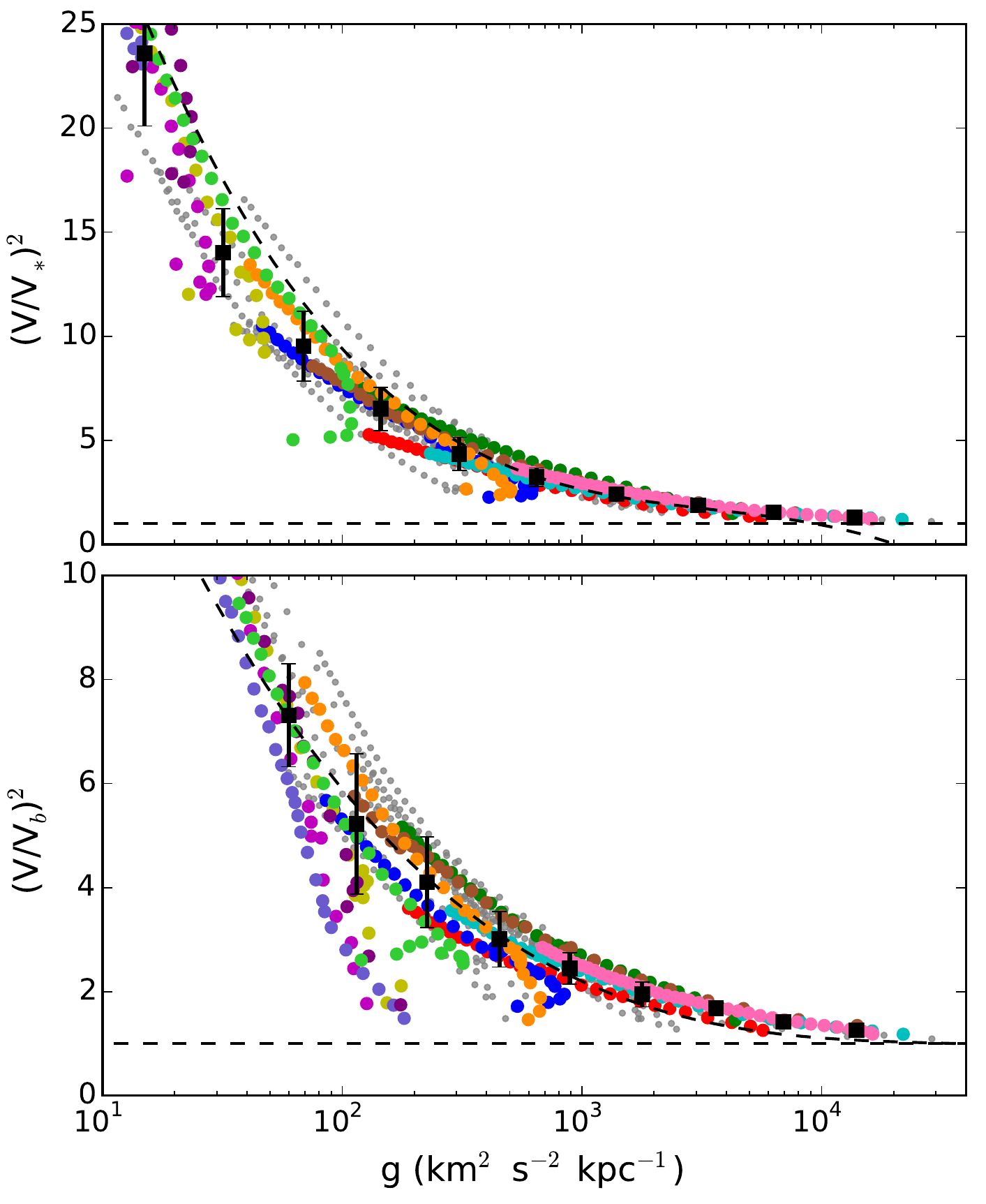}
\caption{Mass discrepancy versus acceleration produced by stars (top panel), and by all baryons--stars + cold gas--(lower panel). The 22 galaxies are shown with colors as in Figures \ref{fig:magicc_rc} and \ref{fig:clues_rc}. The dashed lines are the observational D-g relations (equations 8 \& 9 of \citealt{mcgaugh14}). An horizontal dashed line is also shown in both panels to emphasize the asymptotic behaviour of the relation to D=1. Binned data is shown as black squares. The errors are the \textit{rms} deviations from the best 3 degree polynomial fit found for the data in each bin.
}
\label{fig:Dvsg}
\end{figure}

\subsection{Mass discrepancy}
``Mass discrepancy" refers to the difference 
between the total mass and the baryonic mass enclosed at a certain radius, which can be inferred from the rotation curve of a galaxy. 
The value of the mass discrepancy D is calculated as the squared ratio of the observed velocity to that due to the observed baryons D=(V$_{r}$/V$_{\rm b}$)$^2$.
 
Figure \ref{fig:DvsR} shows mass discrepancy plotted against radius, each point representing a  point along the rotation curves (MaGICC galaxies are colored points, CLUES are small gray dots). 
In the upper panel mass discrepancy is computed as the squared ratio of the observed velocity to that predicted by the stars, and in the lower panel all baryons (stars plus cold gas) are taken into account.

At the radii where D$=$1, the rotation of the galaxy can be explained by the contribution of baryons (stars in the upper panel) alone, while the mass discrepancy (need for dark matter) appears when D$>$1. As with observed galaxies, the mass discrepancy  does not  appear at the same radius for all galaxies, and increases with radius but not at a constant rate for every system \citep{mcgaugh14}.  
Furthermore, galaxies separate readily when mass discrepancy is plotted against radius, with low mass galaxies having a larger  dark matter contribution to the mass at any given radius, compared to higher mass galaxies. 

We note the most prominent difference between  simulations  and observations for the relations shown in Figure \ref{fig:DvsR} is that (V/V$_b$)$^2$ is higher at low radii  for the lowest mass observations, than for the corresponding data from the simulations. As this difference does not appear in the (V/V$_*$)$^2$ case, one may infer that the simulations have an excess of cold gas in the inner regions of low mass galaxies.

Figure \ref{fig:Dvsg} shows the mass discrepancy-acceleration relation for all the simulated galaxies, where 
 the acceleration  is derived from the gravitational potential of the stars (top panel) and baryons (lower panel), g$\equiv$V$^2$/r.
  In this plot, since radius is inversely proportional to the gravitational acceleration, an increase in radius along a rotation curve is read from right to left. One can observe that more massive galaxies reach higher values of g. Note that although
 individual galaxies inhabit different regions of the plot, as can be readily seen for the MaGICC set which are plotted as different colors, they all follow a single relation. 
  
We divide the data in 10 bins and show in Figure  \ref{fig:Dvsg} the fit to the data in each bin as black squares. Error bars show the standard deviation within each bin. 
Dashed lines represent the fits found for observational data, using equations 8 \& 9 of \citet{mcgaugh14}.

A slight deviation from the general trend can be seen at g$\sim$10$^2$ and D$\sim$2, with some points of the simulated galaxies falling below the D-g relation. This feature is also seen in the observed mass discrepancy--acceleration relations. In the simulations, it is specially evident for MaGICC galaxies g21647\_Irr, g7124\_Irr, g5664\_Irr and g1536\_Irr in the ``all baryons" relation.  These are  the galaxies that have the most slowly rising rotation curves, as mentioned in section \ref{rcs}. As discussed above, the stronger feedback scheme present in these simulations removes more baryonic mass from the centre of some of the simulated galaxies, causing the velocity in this inner region to be lower than expected by the one-to-one D-g relation.    Better statistics are required to determine whether the deviations from the relation seen  in the simulations are more or less prominent than seen in the observations, and may relate to the observed diversity of rotation curve shapes. 
 
Another possible difference between the observations and simulations is found in the baryonic D-g relation, where the observed galaxies extend down to D=1, while the simulated galaxies do not quite reach this asymptotic value. Comparing such fine details of the relations would likely require better matched observational and simulated data, in terms of the distributions of stellar masses and scale-lengths.  In particular, the observed sample appears to have more massive disc galaxies, which may be dominated by baryons in the inner region and hence extend down to lower values of D than the simulated sample, which has only 1 galaxy with \mstar$>$3$\times$10$^{10}$\msun. 
 
\begin{figure}
\includegraphics[width=\linewidth]{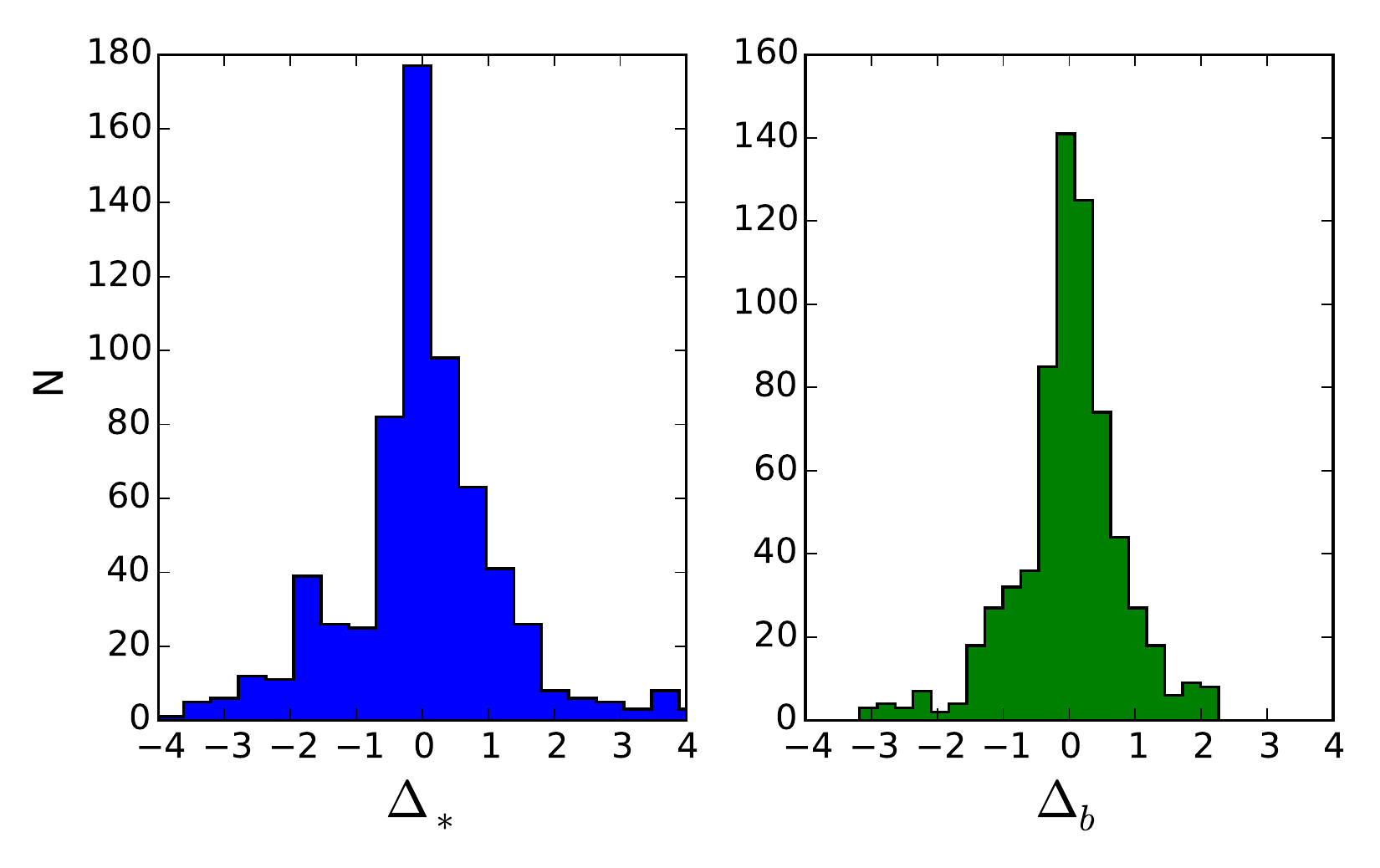}
\caption{Histograms of the residuals around the fits. Left: only stellar contribution; Right: all baryons. }
\label{fig:residuals}
\end{figure}

 The intrinsic scatter we find with respect to the fit in each bin, as well as the deviation from the observed relations, is small and decreases as D tends to 1.
In Figure \ref{fig:residuals} the histograms of the residuals found around the fits are shown to be narrow, with a very similar spread to that of observed galaxies. 
 
\section{Conclusions}
In this paper we have moved beyond using the Tully-Fisher relation as a test of the angular momentum content of galaxies simulated in a \lcdm\ cosmological context, to include the wealth of information contained within extended rotation curve data. We do this by exploring the mass discrepancy-acceleration relation through the full radial range  of the disc in each of 22 simulated galaxies,  a  suite  that spans  more than two orders of magnitude in stellar mass, with rotation velocities ranging from 52 to 222 \kms. 

The simulated suite of late type galaxies is shown to match the empirical relation between baryonic mass, which includes their stars and cold gas, and halo mass, as well as the  baryonic Tully-Fisher relation.

Despite showing significant diversity in the shapes of their rotation curves,  and in the contribution of dark matter to the total mass budget, the simulated galaxies follow a single relation in both the D-g$_*$ and D-g$_b$ plots, with small scatter.  The implication is that not only the total amount of angular momentum attained by the simulated galaxies is correct,  as shown by the BTFR,  but also that their final internal distribution of  gas, stars and dark matter at all radii through their discs is similar to that observed in real galaxies.

The acquisition of the angular momentum within the simulations is complicated, as compared to the simple models of angular momentum acquisition within a CDM  universe that assume collapsing spheres of gas, torqued by large scale structure \citep{fall80}. The baryon cycle within the simulated discs studied here \citep[see][for details]{brook14} involve  a complex web structure, large scale outflows of low angular momentum gas \citep{brook11}, and redistribution of low angular momentum gas through large scale galactic fountains  \citep{brook12a}.
Further, the  dark matter distributions respond  to the gas flows in a manner that is dependent on the simulated galaxy mass \citep{DiCintio2014a, DiCintio2014b}. 

Yet in many ways, our model remains straight forward once the \lcdm\ initial conditions are set, with much of the physics involved in driving the complicated galactic scale baryon cycle occurring on "sub-grid" scales, and accounted for by using relatively simple prescriptions. These prescriptions have improved markedly over the past decade, such that we are able to simulate populations of galaxies that have global properties similar to observed galaxy populations. Our study shows that the distribution of mass within  disc  galaxies with a wide range of masses and rotational velocities can now also be well reproduced within a \lcdm\ universe.

\section*{Acknowledgements}
We thank Yehuda Hoffman for making the CLUES initial conditions available.
ISS, CB \& GY thank the MINECO (Spain) for the financial support through the AYA2012-31101 grant.
CB is funded through the Ramon y Cajal program of the MINECO. 
The CLUES simulations were performed and analyzed at the High Performance Computing Center Stuttgart (HLRS). 
We thank DEISA for  access to  computing resources  through DECI projects SIMU-LU and SIMUGAL-LU and the generous allocation of resources
from STFC’s DiRAC Facility (COSMOS: Galactic Archaeology),
the DEISA consortium, co-funded through EU FP6
project RI-031513 and the FP7 project RI-222919 (through the
DEISA Extreme Computing Initiative), the PRACE-2IP Project (FP7 RI-283493).


\bibliographystyle{mn2e}
\bibliography{MassDis_paper}

\begin{thebibliography}{66}
\expandafter\ifx\csname natexlab\endcsname\relax\def\natexlab#1{#1}\fi

\bibitem[{{Aumer} {et~al}\mbox{.}(2013){Aumer}, {White}, {Naab}, \&
  {Scannapieco}}]{aumer13}
{Aumer} M., {White} S.~D.~M., {Naab} T., {Scannapieco} C., 2013, \mnras, 434,
  3142

\bibitem[{{Barnes} \& {Efstathiou}(1987)}]{barnes87}
{Barnes} J., {Efstathiou} G., 1987, ApJ, 319, 575

\bibitem[{{Begeman} {et~al}\mbox{.}(1991){Begeman}, {Broeils}, \&
  {Sanders}}]{begeman91}
{Begeman} K.~G., {Broeils} A.~H., {Sanders} R.~H., 1991, \mnras, 249, 523

\bibitem[{{Brook}(2015)}]{brook15a}
{Brook} C., 2015, arXiv.1506.00214

\bibitem[{{Brook} {et~al}\mbox{.}(2011){Brook}, {Governato}, {Ro{\v s}kar},
  {Stinson}, {Brooks}, {Wadsley}, {Quinn}, {Gibson}, {Snaith}, {Pilkington},
  {House}, \& {Pontzen}}]{brook11}
{Brook} C.~B. {et~al.}, 2011, MNRAS, 595

\bibitem[{{Brook} {et~al}\mbox{.}(2012{\natexlab{a}}){Brook}, {Stinson},
  {Gibson}, {Ro{\v s}kar}, {Wadsley}, \& {Quinn}}]{brook12a}
{Brook} C.~B., {Stinson} G., {Gibson} B.~K., {Ro{\v s}kar} R., {Wadsley} J.,
  {Quinn} T., 2012{\natexlab{a}}, MNRAS, 419, 771

\bibitem[{{Brook} {et~al}\mbox{.}(2014){Brook}, {Stinson}, {Gibson}, {Shen},
  {Macci{\`o}}, {Obreja}, {Wadsley}, \& {Quinn}}]{brook14}
{Brook} C.~B., {Stinson} G., {Gibson} B.~K., {Shen} S., {Macci{\`o}} A.~V.,
  {Obreja} A., {Wadsley} J., {Quinn} T., 2014, \mnras, 443, 3809

\bibitem[{{Brook} {et~al}\mbox{.}(2012{\natexlab{b}}){Brook}, {Stinson},
  {Gibson}, {Wadsley}, \& {Quinn}}]{brook12}
{Brook} C.~B., {Stinson} G., {Gibson} B.~K., {Wadsley} J., {Quinn} T.,
  2012{\natexlab{b}}, MNRAS, 424, 1275

\bibitem[{{Cantalupo}(2010)}]{cantalupo10}
{Cantalupo} S., 2010, \mnras, 403, L16

\bibitem[{{de Blok} {et~al}\mbox{.}(2001){de Blok}, {McGaugh}, {Bosma}, \&
  {Rubin}}]{deblok01}
{de Blok} W.~J.~G., {McGaugh} S.~S., {Bosma} A., {Rubin} V.~C., 2001, \apjl,
  552, L23

\bibitem[{{Di Cintio} {et~al}\mbox{.}(2014{\natexlab{a}}){Di Cintio}, {Brook},
  {Dutton}, {Macci{\`o}}, {Stinson}, \& {Knebe}}]{DiCintio2014b}
{Di Cintio} A., {Brook} C.~B., {Dutton} A.~A., {Macci{\`o}} A.~V., {Stinson}
  G.~S., {Knebe} A., 2014{\natexlab{a}}, \mnras, 441, 2986

\bibitem[{{Di Cintio} {et~al}\mbox{.}(2014{\natexlab{b}}){Di Cintio}, {Brook},
  {Macci{\`o}}, {Stinson}, {Knebe}, {Dutton}, \& {Wadsley}}]{DiCintio2014a}
{Di Cintio} A., {Brook} C.~B., {Macci{\`o}} A.~V., {Stinson} G.~S., {Knebe} A.,
  {Dutton} A.~A., {Wadsley} J., 2014{\natexlab{b}}, \mnras, 437, 415

\bibitem[{{Dom{\'e}nech-Moral} {et~al}\mbox{.}(2012){Dom{\'e}nech-Moral},
  {Mart{\'{\i}}nez-Serrano}, {Dom{\'{\i}}nguez-Tenreiro}, \&
  {Serna}}]{domenech12}
{Dom{\'e}nech-Moral} M., {Mart{\'{\i}}nez-Serrano} F.~J.,
  {Dom{\'{\i}}nguez-Tenreiro} R., {Serna} A., 2012, \mnras, 421, 2510

\bibitem[{{Dom{\'{\i}}nguez-Tenreiro}
  {et~al}\mbox{.}(2015){Dom{\'{\i}}nguez-Tenreiro}, {Obreja}, {Brook},
  {Mart{\'{\i}}nez-Serrano}, {Stinson}, \& {Serna}}]{dominguez15}
{Dom{\'{\i}}nguez-Tenreiro} R., {Obreja} A., {Brook} C.~B.,
  {Mart{\'{\i}}nez-Serrano} F.~J., {Stinson} G., {Serna} A., 2015, \apjl, 800,
  L30

\bibitem[{{Fall} \& {Efstathiou}(1980)}]{fall80}
{Fall} S.~M., {Efstathiou} G., 1980, MNRAS, 193, 189

\bibitem[{{Feldmann} \& {Mayer}(2015)}]{feldman15}
{Feldmann} R., {Mayer} L., 2015, \mnras, 446, 1939

\bibitem[{{Gentile} {et~al}\mbox{.}(2004){Gentile}, {Salucci}, {Klein},
  {Vergani}, \& {Kalberla}}]{gentile04}
{Gentile} G., {Salucci} P., {Klein} U., {Vergani} D., {Kalberla} P., 2004,
  \mnras, 351, 903

\bibitem[{{Gottloeber} {et~al}\mbox{.}(2010){Gottloeber}, {Hoffman}, \&
  {Yepes}}]{gottloeber10}
{Gottloeber} S., {Hoffman} Y., {Yepes} G., 2010, arXiv.1005.2687

\bibitem[{{Governato} {et~al}\mbox{.}(2010){Governato}, {Brook}, {Mayer},
  {Brooks}, {Rhee}, {Wadsley}, {Jonsson}, {Willman}, {Stinson}, {Quinn}, \&
  {Madau}}]{governato10}
{Governato} F. {et~al.}, 2010, Nature, 463, 203

\bibitem[{{Governato} {et~al}\mbox{.}(2004){Governato}, {Mayer}, {Wadsley},
  {Gardner}, {Willman}, {Hayashi}, {Quinn}, {Stadel}, \& {Lake}}]{governato04}
{Governato} F. {et~al.}, 2004, \apj, 607, 688

\bibitem[{{Guedes} {et~al}\mbox{.}(2011){Guedes}, {Callegari}, {Madau}, \&
  {Mayer}}]{guedes11}
{Guedes} J., {Callegari} S., {Madau} P., {Mayer} L., 2011, \apj, 742, 76

\bibitem[{{Guo} {et~al}\mbox{.}(2010){Guo}, {White}, {Li}, \&
  {Boylan-Kolchin}}]{guo10}
{Guo} Q., {White} S., {Li} C., {Boylan-Kolchin} M., 2010, MNRAS, 404, 1111

\bibitem[{{Hopkins} {et~al}\mbox{.}(2014){Hopkins}, {Kere{\v s}}, {O{\~n}orbe},
  {Faucher-Gigu{\`e}re}, {Quataert}, {Murray}, \& {Bullock}}]{hopkins14}
{Hopkins} P.~F., {Kere{\v s}} D., {O{\~n}orbe} J., {Faucher-Gigu{\`e}re} C.-A.,
  {Quataert} E., {Murray} N., {Bullock} J.~S., 2014, \mnras, 445, 581

\bibitem[{{Kannan} {et~al}\mbox{.}(2013){Kannan}, {Stinson}, {Macci{\`o}},
  {Brook}, {Weinmann}, {Wadsley}, \& {Couchman}}]{kannan13}
{Kannan} R., {Stinson} G.~S., {Macci{\`o}} A.~V., {Brook} C., {Weinmann} S.~M.,
  {Wadsley} J., {Couchman} H.~M.~P., 2013, arXiv.1302.2618

\bibitem[{{Kannan} {et~al}\mbox{.}(2014){Kannan}, {Stinson}, {Macci{\`o}},
  {Hennawi}, {Woods}, {Wadsley}, {Shen}, {Robitaille}, {Cantalupo}, {Quinn}, \&
  {Christensen}}]{Kannan14b}
{Kannan} R. {et~al.}, 2014, \mnras, 437, 2882

\bibitem[{{Knollmann} \& {Knebe}(2009)}]{knollmann09}
{Knollmann} S.~R., {Knebe} A., 2009, \apjs, 182, 608

\bibitem[{{Kuzio de Naray} {et~al}\mbox{.}(2006){Kuzio de Naray}, {McGaugh},
  {de Blok}, \& {Bosma}}]{kuziodenaray06}
{Kuzio de Naray} R., {McGaugh} S.~S., {de Blok} W.~J.~G., {Bosma} A., 2006,
  \apjs, 165, 461

\bibitem[{{Maller} \& {Dekel}(2002)}]{maller02}
{Maller} A.~H., {Dekel} A., 2002, MNRAS, 335, 487

\bibitem[{{McGaugh}(2014)}]{mcgaugh14}
{McGaugh} S., 2014, Galaxies, 2, 601

\bibitem[{{McGaugh}(2004)}]{mcgaugh04}
{McGaugh} S.~S., 2004, \apj, 609, 652

\bibitem[{{McGaugh}(2005)}]{mcgaugh05}
{McGaugh} S.~S., 2005, \apj, 632, 859

\bibitem[{{McGaugh}(2012)}]{mcgaugh12}
{McGaugh} S.~S., 2012, \aj, 143, 40

\bibitem[{{McGaugh} \& {Schombert}(2015)}]{McGaugh15}
{McGaugh} S.~S., {Schombert} J.~M., 2015, \apj, 802, 18

\bibitem[{{Milgrom}(1983)}]{milgrom83}
{Milgrom} M., 1983, \apj, 270, 365

\bibitem[{{Moster} {et~al}\mbox{.}(2013){Moster}, {Naab}, \&
  {White}}]{moster13}
{Moster} B.~P., {Naab} T., {White} S.~D.~M., 2013, MNRAS, 428, 3121

\bibitem[{{Moster} {et~al}\mbox{.}(2010){Moster}, {Somerville}, {Maulbetsch},
  {van den Bosch}, {Macci{\`o}}, {Naab}, \& {Oser}}]{moster10}
{Moster} B.~P., {Somerville} R.~S., {Maulbetsch} C., {van den Bosch} F.~C.,
  {Macci{\`o}} A.~V., {Naab} T., {Oser} L., 2010, ApJ, 710, 903

\bibitem[{{Munshi} {et~al}\mbox{.}(2013){Munshi}, {Governato}, {Brooks},
  {Christensen}, {Shen}, {Loebman}, {Moster}, {Quinn}, \& {Wadsley}}]{munshi13}
{Munshi} F. {et~al.}, 2013, \apj, 766, 56

\bibitem[{{Navarro} \& {Steinmetz}(2000)}]{navarrosteinmetz00}
{Navarro} J.~F., {Steinmetz} M., 2000, ApJ, 538, 477

\bibitem[{{Obreja} {et~al}\mbox{.}(2014){Obreja}, {Brook}, {Stinson},
  {Dom{\'{\i}}nguez-Tenreiro}, {Gibson}, {Silva}, \& {Granato}}]{obreja14}
{Obreja} A., {Brook} C.~B., {Stinson} G., {Dom{\'{\i}}nguez-Tenreiro} R.,
  {Gibson} B.~K., {Silva} L., {Granato} G.~L., 2014, \mnras, 442, 1794

\bibitem[{{Oh} {et~al}\mbox{.}(2015){Oh}, {Hunter}, {Brinks}, {Elmegreen},
  {Schruba}, {Walter}, {Rupen}, {Young}, {Simpson}, {Johnson}, {Herrmann},
  {Ficut-Vicas}, {Cigan}, {Heesen}, {Ashley}, \& {Zhang}}]{oh15}
{Oh} S.-H. {et~al.}, 2015, arXiv.1502.01281

\bibitem[{{Oman} {et~al}\mbox{.}(2015){Oman}, {Navarro}, {Fattahi}, {Frenk},
  {Sawala}, {White}, {Bower}, {Crain}, {Furlong}, {Schaller}, {Schaye}, \&
  {Theuns}}]{oman15}
{Oman} K.~A. {et~al.}, 2015, arXiv.1504.01437

\bibitem[{{Papastergis} {et~al}\mbox{.}(2012){Papastergis}, {Cattaneo},
  {Huang}, {Giovanelli}, \& {Haynes}}]{papastergis12}
{Papastergis} E., {Cattaneo} A., {Huang} S., {Giovanelli} R., {Haynes} M.~P.,
  2012, \apj, 759, 138

\bibitem[{{Peebles}(1969)}]{peebles69}
{Peebles} P.~J.~E., 1969, ApJ, 155, 393

\bibitem[{{Pichon} {et~al}\mbox{.}(2011){Pichon}, {Pogosyan}, {Kimm}, {Slyz},
  {Devriendt}, \& {Dubois}}]{pichon11}
{Pichon} C., {Pogosyan} D., {Kimm} T., {Slyz} A., {Devriendt} J., {Dubois} Y.,
  2011, \mnras, 418, 2493

\bibitem[{{Piontek} \& {Steinmetz}(2011)}]{piontek11}
{Piontek} F., {Steinmetz} M., 2011, MNRAS, 410, 2625

\bibitem[{{Pontzen} {et~al}\mbox{.}(2013){Pontzen}, {Ro{\v s}kar}, {Stinson},
  {Woods}, {Reed}, {Coles}, \& {Quinn}}]{pynbody}
{Pontzen} A., {Ro{\v s}kar} R., {Stinson} G.~S., {Woods} R., {Reed} D.~M.,
  {Coles} J., {Quinn} T.~R., 2013, {pynbody: Astrophysics Simulation Analysis
  for Python}. Astrophysics Source Code Library, ascl:1305.002

\bibitem[{{Prochaska} {et~al}\mbox{.}(2011){Prochaska}, {Weiner}, {Chen},
  {Mulchaey}, \& {Cooksey}}]{prochaska11}
{Prochaska} J.~X., {Weiner} B., {Chen} H.-W., {Mulchaey} J., {Cooksey} K.,
  2011, ApJ, 740, 91

\bibitem[{{Rubin} \& {Ford}(1970)}]{rubin70}
{Rubin} V.~C., {Ford}, Jr. W.~K., 1970, \apj, 159, 379

\bibitem[{{Sancisi}(2004)}]{renzo04}
{Sancisi} R., 2004, in IAU Symposium, Vol. 220, Dark Matter in Galaxies,
  {Ryder} S., {Pisano} D., {Walker} M., {Freeman} K., eds., p. 233

\bibitem[{{Sanders} \& {McGaugh}(2002)}]{sanders02}
{Sanders} R.~H., {McGaugh} S.~S., 2002, \araa, 40, 263

\bibitem[{{Sanders} \& {Verheijen}(1998)}]{sanders98}
{Sanders} R.~H., {Verheijen} M.~A.~W., 1998, \apj, 503, 97

\bibitem[{{Schaye} {et~al}\mbox{.}(2015){Schaye}, {Crain}, {Bower}, {Furlong},
  {Schaller}, {Theuns}, {Dalla Vecchia}, {Frenk}, {McCarthy}, {Helly},
  {Jenkins}, {Rosas-Guevara}, {White}, {Baes}, {Booth}, {Camps}, {Navarro},
  {Qu}, {Rahmati}, {Sawala}, {Thomas}, \& {Trayford}}]{schaye15}
{Schaye} J. {et~al.}, 2015, \mnras, 446, 521

\bibitem[{{Shen} {et~al}\mbox{.}(2010){Shen}, {Wadsley}, \& {Stinson}}]{shen10}
{Shen} S., {Wadsley} J., {Stinson} G., 2010, \mnras, 407, 1581

\bibitem[{{Stinson} {et~al}\mbox{.}(2006){Stinson}, {Seth}, {Katz}, {Wadsley},
  {Governato}, \& {Quinn}}]{stinson06}
{Stinson} G., {Seth} A., {Katz} N., {Wadsley} J., {Governato} F., {Quinn} T.,
  2006, \mnras, 373, 1074

\bibitem[{{Stinson} {et~al}\mbox{.}(2010){Stinson}, {Bailin}, {Couchman},
  {Wadsley}, {Shen}, {Nickerson}, {Brook}, \& {Quinn}}]{stinson10}
{Stinson} G.~S., {Bailin} J., {Couchman} H., {Wadsley} J., {Shen} S.,
  {Nickerson} S., {Brook} C., {Quinn} T., 2010, \mnras, 408, 812

\bibitem[{{Stinson} {et~al}\mbox{.}(2013){Stinson}, {Brook}, {Macci{\`o}},
  {Wadsley}, {Quinn}, \& {Couchman}}]{stinson13}
{Stinson} G.~S., {Brook} C., {Macci{\`o}} A.~V., {Wadsley} J., {Quinn} T.~R.,
  {Couchman} H.~M.~P., 2013, MNRAS, 428, 129

\bibitem[{{Stinson} {et~al}\mbox{.}(2012){Stinson}, {Brook}, {Prochaska},
  {Hennawi}, {Shen}, {Wadsley}, {Pontzen}, {Couchman}, {Quinn}, {Macci{\`o}},
  \& {Gibson}}]{stinson12}
{Stinson} G.~S. {et~al.}, 2012, MNRAS, 425, 1270

\bibitem[{{Swaters} {et~al}\mbox{.}(2009){Swaters}, {Sancisi}, {van Albada}, \&
  {van der Hulst}}]{swaters09}
{Swaters} R.~A., {Sancisi} R., {van Albada} T.~S., {van der Hulst} J.~M., 2009,
  \aap, 493, 871

\bibitem[{{Thacker} \& {Couchman}(2001)}]{thacker01}
{Thacker} R.~J., {Couchman} H.~M.~P., 2001, ApJL, 555, L17

\bibitem[{{Tully} \& {Verheijen}(1997)}]{tully97}
{Tully} R.~B., {Verheijen} M.~A.~W., 1997, \apj, 484, 145

\bibitem[{{Tumlinson} {et~al}\mbox{.}(2011){Tumlinson}, {Thom}, {Werk},
  {Prochaska}, {Tripp}, {Weinberg}, {Peeples}, {OMeara}, {Oppenheimer},
  {Meiring}, {Katz}, {Dav{\'e}}, {Ford}, \& {Sembach}}]{tumlinson11}
{Tumlinson} J. {et~al.}, 2011, Science, 334, 948

\bibitem[{{van den Bosch} \& {Dalcanton}(2000)}]{vdb00}
{van den Bosch} F.~C., {Dalcanton} J.~J., 2000, \apj, 534, 146

\bibitem[{{Wadsley} {et~al}\mbox{.}(2004){Wadsley}, {Stadel}, \&
  {Quinn}}]{wadsley04}
{Wadsley} J.~W., {Stadel} J., {Quinn} T., 2004, \na, 9, 137

\bibitem[{{Wang} {et~al}\mbox{.}(2015){Wang}, {Dutton}, {Stinson},
  {Macci{\`o}}, {Penzo}, {Kang}, {Keller}, \& {Wadsley}}]{wang15}
{Wang} L., {Dutton} A.~A., {Stinson} G.~S., {Macci{\`o}} A.~V., {Penzo} C.,
  {Kang} X., {Keller} B.~W., {Wadsley} J., 2015, arXiv.1503.04818

\bibitem[{{Yepes} {et~al}\mbox{.}(2014){Yepes}, {Gottl{\"o}ber}, \&
  {Hoffman}}]{yepes14}
{Yepes} G., {Gottl{\"o}ber} S., {Hoffman} Y., 2014, \nar, 58, 1

\bibitem[{{Zwaan} {et~al}\mbox{.}(1995){Zwaan}, {van der Hulst}, {de Blok}, \&
  {McGaugh}}]{zwaan95}
{Zwaan} M.~A., {van der Hulst} J.~M., {de Blok} W.~J.~G., {McGaugh} S.~S.,
  1995, \mnras, 273, L35

\end{thebibliography}


\label{lastpage}

\end{document}